# An electronic radiation of blackbody: Cosmic electron background

Jian-Miin Liu (刘建民)*
Department of Physics, Nanjing University
Nanjing, The People's Republic of China
*On leave. E-mail: liu@phys.uri.edu

Abstract

The Universe owns the electronic radiation of blackbody at temperature 2.725 K, which we call the cosmic electron background. We calculate its radiation spectrum. The energy distribution of number density of electrons in the cosmic electron background becomes zero as energy goes to both zero and infinity. It has one maximum peak near the energy level of $10^{-23}$ J.

The spectrum of cosmic microwave background was firstly discovered by Penzias and Wilson in 1972 [1]. Later, it was confirmed by different research groups, especially those groups using the "Cosmic Background Explorer" (COBE) satellite system launched in November of 1989, to be a one of radiation of nearly perfect blackbody at temperature

$T_u$=2.725 K [2-7].

Together, it became clear that the cosmic microwave background is almost isotropic with an anisotropy at the level of a part of 100,000. The existence of such cosmic microwave background indicates that the Universe can be at least theoretically counted to be a cavity acting as a nearly perfect blackbody, in which all electromagnetic radiations or all photons emitted by various sources have been in thermal equilibrium with each other at $T_u$, because the thermo-equilibrium radiation inside a cavity being at temperature $T$ has the same character as the radiation emitted by a blackbody at temperature $T$. We might call it the Universe blackbody-cavity sometimes.

Another fact to note is that there are stars of a low density in the Universe and many of them house atomic transitions, nuclear fusion reactions and other high-energy particle reactions. In doing, these stars emit not only electromagnetic radiations but also particle radiations, like electrons.

Now let us assume that in the Universe blackbody-cavity all electronic radiations or all electrons emitted by these stars have been also in thermal equilibrium with each other [8]. According to this assumption, the Universe would own the electronic radiation of blackbody at temperature $T_u$ that we call the cosmic electron background. In this letter, we calculate its radiation spectrum.

The radiation of a blackbody is a thermo-equilibrium radiation. Its spectrum depends only on the temperature of the blackbody. For the electronic radiation of a blackbody at low temperature, we can use the Fermi-Dirac quantum equilibrium distribution [9], $F = 1/[\exp(\frac{E-\mu}{K_B T})+1]$, to calculate its spectrum, where $F$ is the probable number of electrons per quantum state, $E$ the energy of quantum state, $\mu$ the chemical potential, $K_B$ the Boltzmann constant and $T$ the temperature of the blackbody. Taking into account two possible orientations for an electron in volume $V$, one can find that the number of states in the energy interval, from $E$ to $E+dE$, is $\frac{4\pi V}{h^3}(2m_e)^{3/2} E^{1/2} dE$, where $m_e$ is the rest mass of electrons and $h$ the Planck constant. The energy distribution of number density of electrons in the Universe blackbody-cavity is therefore

$$n(E)dE \equiv \frac{N(E)}{V}dE = \frac{4\pi}{h^3}(2m_e)^{3/2} \frac{1}{\exp(\frac{E-\mu_u}{K_B T_u})+1} E^{1/2} dE, \tag{1}$$

where $N(E)dE$ is the energy distribution of number of electrons in volume $V$ in the Universe and $\mu_u$ the chemical potential of electrons at temperature $T_u$. This distribution satisfies

$$\int_0^\infty n(E)dE = \int_0^\infty \frac{4\pi}{h^3}(2m_e)^{3/2} \frac{1}{\exp(\frac{E-\mu_u}{K_B T_u})+1} E^{1/2} dE = n_e, \tag{2}$$

where $n_e$ is the total number density of electrons in the Universe. By Eq.(2), chemical potential $\mu_u$ is connected to temperature $T_u$ and number density $n_e$.

Chemical potential varies with temperature, but very slowly at low temperature [10,11]. For electrons in the cosmic electron background, chemical potential $\mu_u$ at $T_u$ has the same order as that at zero temperature. It is not hard to evaluate the zero-temperature chemical potential, denoted with $\mu_0$. Since, at zero temperature, the Fermi-Dirac distribution is

$F= 0$, for $E > \mu_0$;
$F= 1$, for $E < \mu_0$,

Eq.(2) becomes

$$\int_0^{\mu_0} \frac{4\pi}{h^3}(2m_e)^{3/2} E^{1/2} dE = n_e.$$

Completing the integral gives rise to

$$\mu_0 = \frac{\eta^2}{2m_e}(3\pi^2 n_e)^{2/3}, \tag{3}$$

which is proportional to $(n_e)^{2/3}$, where $\eta = h/2\pi$. The Planck constant $\eta$ is of the order of $10^{-34}$ and $m_e$ is of the order of $10^{-31}$, so the order of chemical potential $\mu_0$ will be less than $10^{-27}$ J provided the order of $n_e$ is less than $10^{15}$ m$^{-3}$. Such chemical potential is very small compared to

$$K_B T_u \approx 10^{-23} J.$$

Actually, the total number density of electrons in the Universe, $n_e$, is in general estimated to be of the order of $10^{-1}$ m$^{-3}$ [12,13], much smaller than $10^{15}$ m$^{-3}$. If we take this estimate, we learn that $\mu_u \approx \mu_0$ is of the order of $10^{-36}$ J and, hence, $\mu_u / K_B T_u \approx 10^{-13}$ which is small enough to be neglected from Eq.(1). We finally obtain

$$n(E)dE = \frac{4\pi}{h^3}(2m_e)^{3/2}\frac{1}{\exp(\frac{E}{K_B T_u})+1}E^{1/2}dE, \qquad (4)$$

as the energy distribution of number density of electrons in the Universe, namely the radiation spectrum of cosmic electron background.

The distribution function, $n(E)$, vanishes as energy $E$ goes to zero or infinity. It has only one maximum peak. To prove this, we take its derivative with respect to energy $E$ and get

$$n'(E) = \frac{4\pi}{h^3}\frac{(2m_e)^{3/2}}{2E^{1/2}(e^{E/K_B T_u}+1)^2}[(1-\frac{2E}{K_B T_u})e^{E/K_B T_u}+1]. \qquad (5)$$

We rewrite Eq.(5) as

$$n'(E) = (\text{a positive})[(1-\frac{2E}{K_B T_u})e^{E/K_B T_u}+1] \qquad (6)$$

for simplification. Putting $E=K_B T_u$ and $E=K_B T_u/2$ into Eq.(6) respectively, we find

$n'(K_B T_u / 2) > 0$ and $n'(K_B T_u) < 0$.

Thus, there is at least one energy value between $K_B T_u/2$ and $K_B T_u$, denoted with $E_m$, that makes $n'(E_m)$ equal zero, i.e.

$$n'(E_m) = 0. \qquad (7)$$

If you plot two functions, $n'_1(E) = 1-\frac{2E}{K_B T_u}$ and $n'_2(E) = e^{E/K_B T_u}$, on the plane of $n'(E)$ versus $E$, $E>0$, you can see that $n'_1(E)$ monotonously decreases with $E$ and

$n'_1(E) > 0$ for $E < K_B T_u / 2$;
$n'_1(E) < 0$ for $E > K_B T_u / 2$.

You can also see that $n'_2(E)$ monotonously increases with $E$ and

$n'_2(E) > 1$ for $E > 0$.

So, it must be that

$n'_1(E)n'_2(E) < -1$, for $E > E_m$;
$0 > n'_1(E)n'_2(E) > -1$, for $K_B T_u / 2 < E < E_m$;
$n'_1(E)n'_2(E) > 0$, for $0 < E < K_B T_u / 2$,

because $n'_1(E_m)n'_2(E_m)$ equals $-1$ due to Eqs.(6) and (7). Furthermore, because $n'(E) = (\text{a positive})[n'_1(E)n'_2(E)+1]$, we immediately have

$n'(E) < 0$, for $E > E_m$;  (8a)

$n'(E) > 0$, for $E < E_m$.  (8b)

Inequalities (8a-b) combine with Eq.(7) to ensure that $n'(E)$ does not have any zero-point other than $E_m$. In other words, distribution function $n(E)$ has only one maximum peak at $E_m$.

In sum, (1) the energy distribution of number density of electrons in the cosmic electron background becomes zero as energy $E$ goes to both zero and infinity; (2) It has one maximum peak at $E_m$; (3) $E_m$ falls on between $K_BT_u/2$ and $K_BT_u$, having the same order as $K_BT_u \approx 10^{-23} J$. That means that most of electrons in the cosmic electron background are distributed near the energy level of $10^{-23}$ J.

The above calculations are valid for other kinds of massive particles obeying the Fermi-Dirac distribution, like protons or neutrons. If all massive particles of a kind emitted by various sources in the Universe have been in their thermal equilibrium, since the rest mass of massive particles of the kind is greater than that of electrons, the zero-temperature chemical potential for massive particles of the kind must be smaller than that for electrons. We are then able to drop $\exp(-\mu_u/K_BT_u)$ in Eq.(1) when the total number density of massive particles of the kind in the Universe is less than $10^{15}$ m$^{-3}$. The energy distribution represented in Eq.(4) therefore remains for the cosmic background of massive particles of the kind. Since $E_m$, as a zero point of function $n'(E)$ in Eq.(6), is irrelevant to the rest mass of massive particles of the kind, this $E_m$ remains, too, for the cosmic background of massive particles of the kind. That is an indication that the so-called cosmic massive-particle sea exists in the Universe, in which the massive particles of each kind are distributed in a similar manner described by Eq.(4) with their different rest mass. Especially, most of these massive particles of each kind crowd near the energy level of $10^{-23} J$.

Acknowledgment
The author greatly appreciates the teachings of Prof. Wo-Te Sen. The author's thanks also go to Dr. Mario Iorio for his help.